\documentclass[10pt]{article}

\usepackage{amsmath}
\usepackage{amssymb}

\usepackage{graphicx}

\usepackage{cite}

\usepackage{color} 

\usepackage[labelfont=bf,labelsep=period,justification=raggedright]{caption}

\makeatletter
\renewcommand{\@biblabel}[1]{\quad#1.}
\makeatother

\date{}

\begin{document}

\begin{flushleft}
{\Large
\textbf{Fat-tailed fluctuations in the size of organizations: the role of social influence}
}
\\
Hernan Mondani$^{1,\ast}$, 
Petter Holme$^{2,3,1,4}$, 
Fredrik Liljeros$^{1,4}$
\\
\bf{1} Department of Sociology, Stockholm University, 10691 Stockholm, Sweden
\\
\bf{2} Department of Energy Science, Sungkyunkwan University, 440-746 Suwon, Korea
\\
\bf{3} IceLab, Department of Physics, Ume\aa~University, 90187 Ume\aa, Sweden
\\
\bf{4} Institute for Futures Studies, Box 591, 10131 Stockholm, Sweden
\\
$\ast$ E-mail: hernan.mondani@sociology.su.se
\end{flushleft}

\section*{Abstract}
Organizational growth processes have consistently been shown to exhibit a fatter-than-Gaussian growth-rate distribution in a variety of settings. Long periods of relatively small changes are interrupted by sudden changes in all size scales. This kind of extreme events can have important consequences for the development of biological and socio-economic systems. Existing models do not derive this aggregated pattern from agent actions at the micro level. We develop an agent-based simulation model on a social network. We take our departure in a model by a Schwarzkopf et al. on a scale-free network. We reproduce the fat-tailed pattern out of internal dynamics alone, and also find that it is robust with respect to network topology. Thus, the social network and the local interactions are a prerequisite for generating the pattern, but not the network topology itself. We further extend the model with a parameter $\delta$ that weights the relative fraction of an individual's neighbours belonging to a given organization, representing a contextual aspect of social influence. In the lower limit of this parameter, the fraction is irrelevant and choice of organization is random. In the upper limit of the parameter, the largest fraction quickly dominates, leading to a winner-takes-all situation. We recover the real pattern as an intermediate case between these two extremes.


\newpage
\section*{Introduction}
The social world is populated by organizations. The organizational landscape is changing all the time, with organizations being created, restructured and dissolved. In this dynamical environment, potentially several thousands of agents interact following different motivations, both at the individual level and at the collective level. The coevolution of various institutional settings, private and public, as well as the existence of actors and activities at different levels of aggregation, makes research on organizational dynamics a complex subject matter.

We study \emph{organizational growth processes}, that is to say, the time evolution of company size for a system of organizations. These processes exhibit statistical regularities, despite their complexity~\cite{AMARA1997}. One main striking regularity concerns the nature of the probability distribution for the growth rate, i.e. how fast the size changes in time. This distribution has two features. First, it follows a fat-tailed pattern, meaning that organizational size changes very little most of the time, but dramatically every once in a while, leading to rare (yet possible) booms and catastrophic crashes. The second feature is that fluctuations (i.e. the variance) in growth rates are less severe for larger organizations, so that the variance in growth is not uniform across all size scales.

These empirical regularities are relevant for several reasons. The large fluctuations observed in real systems are rare, but certainly more likely than if the process were governed by a Gaussian distribution. This kind of behaviour constitutes a counter-intuitive observation, since traditional models (in economics, for example) would expect it to be Gaussian. The fat-tailed pattern adds unpredictability to the system, allowing for extreme events to take place in a short period of time. This has practical consequences for economic development and societal stability. Finally, similar growth statistics have been observed in a wide variety of natural and artificial systems, which makes the understanding of underlying mechanisms behind growth processes an interdisciplinary topic. The reported systems can be mapped along several dimensions, as we illustrate below:

\begin{description}
\item[country:] US~\cite{AMARA1997b,AMARA1997c,AMARA1997,AMARA1998,BOTTA2003,GUPTA2007,PLERO1999b,STANL1996}, Italy~\cite{BOTTA2001,BOTTA2007,BOTTA2002,BOTTA2003b,BOTTA2006,DOSI2005}, Japan~\cite{ISHIK2006,ISHIK2006b,ISHIK2007}, United Kingdom~\cite{HART1996,SINGH1975}, Brazil~\cite{GUPTA2007}, Sweden~\cite{lilje2001},
\item[profitability:] commercial e.g.~\cite{AMARA1997,BOTTA2006,HART1996} and voluntary~\cite{lilje2001} organizations,
\item[industrial sector:] pharmaceutical~\cite{BOTTA2001,DEFAB2003,FU2005,MATIA2004}, furniture~\cite{BOTTA2006}, printing~\cite{BOTTA2003,BOTTA2006}, shoes~\cite{BOTTA2006}, textiles~\cite{BOTTA2002,BOTTA2003b}, metals~\cite{BOTTA2002,BOTTA2003,BOTTA2003}, chemicals~\cite{BOTTA2003}, food~\cite{BOTTA2003},
\item [process:] industrial production~\cite{AMARA1997,BOTTA2002,FAGIO2007,STANL1996}, firm growth of countries in the G7~\cite{GAFFE2003}, investments in mutual funds~\cite{SCHWA2010}, stock price fluctuations~\cite{PLERO1999b}, country Gross Domestic Product (GDP)~\cite{CASTA2009,FAGIO2007,LEE1998}, exports/imports~\cite{PODOB2008}, bird population dynamics~\cite{KEITT2002}, university research output~\cite{PLERO1999}.
\end{description}

Our theoretical approach to the problem is the theory of social mechanisms in the analytical sociology framework~\cite{hedst2005}. In this tradition, models about social phenomena contribute to our theoretical understanding if they make clear the micro-level mechanisms that bring about a certain macro-level outcome, in this case the non-Gaussian growth-rate pattern. It is individuals that, through their actions, bring about macro-level outcomes~\cite{COLEM1986}. The network of individual contacts is the setting in which the actors can influence each other, with activities in economic life being embedded in social networks~\cite{GRANO1985}. Our proposed model has thus a focus on individuals in organizations as the relevant actors, and the network connecting them is of fundamental importance. This approach makes explicit the interplay between individuality and social influence, and shows how it can lead to an unexpected macro outcome.

Within this context, there are at least two possible reasons contributing to the features we observe in the growth-rate pattern. The fat tails could be caused by asymmetries in the structure of the social network that make certain actors more salient or popular than others. Their involvement in the process of influencing group membership would generate these occasional large growth fluctuations. Along this line, the fat-tailed pattern would be a result of the underlying fat-tailed nature of the network structure. Concerning fluctuations being smaller for larger companies, there could be a rich-get-richer phenomenon at play, by which large organizations become larger by a positive feedback process, and thus their size fluctuates less, while small organizations are more sensitive to perturbations caused by member entrance and exit. This is the firm diversification argument~\cite{AMARA1998,BOTTA2001}.

Several models have been proposed over a range of approaches (see overview below), but there is no general consensus about a dominant mechanism to account for the emergence of fat-tailed growth-rate distributions. Additionally, the majority of models in the literature do not focus on the network of agents that are members of an organization, but are rather aggregated models based on economic considerations or on various types of stochastic processes. For this reason, there is a need for models that describe the mechanism by which social actions at the micro level generate the pattern at the macro-level.

We address these two possible reasons by modelling a social network where agents are subject to social influence when it comes deciding on organizational membership. We build and simulate a first model based on by~\cite{SCHWA2010}, the \emph{SAF model}. This first model represents the localized, network-dependent aspect of social influence. Therefore, our first modelling aim is to implement the SAF model and simulate it on different network topologies, in order to explore if a fat-tailed network structure is necessary for observing the non-Gaussian growth pattern. Secondly, we add a context-dependent aspect of social influence. We implement this by in the \emph{extended SAF model} with an influence parameter that weights an individual's membership choice by contextual influence. We propose an alternative to the diversification argument as a possible explanation of fluctuation dependence with organization size, in terms of a combination of network-dependent and context-dependent social influence.

	\subsection*{Overview of existing models}\label{SecOverModStrat}
In this subsection, we describe existing models. None of the existing models, except for the one we take our departure from, incorporate micro-level mechanisms grounded on individuals and their interactions. We classify models into three categories: economic, physical and stochastic. We shall use `group' and `organization' as synonyms from here on.

\emph{Economic models} account for a large fraction of the models of growth processes in organizations. We find economic models for firm sizes as far back as the mid-twentieth century. Simon~\cite{ijiri1977,SIMON1964,SIMON1958} developed the concept of growth opportunities; Sutton later elaborated on this concept. Lucas~\cite{LUCAS1978} considers on the distribution of managerial talent; more recent models also use this notion~\cite{GUPTA2003}. Jovanovic introduced a model for firm learning~\cite{JOVAN1982} through an evolutionary-like process. In more recent times, Amaral et al. built a model on the concept of optimal size~\cite{AMARA1997c,AMARA1997}. A study by Bottazzi~\cite{BOTTA2001} used the concept of market diversification. Finally, Dosi studied the relation of growth with innovation and production efficiency~\cite{DOSI2005}.

Another model category is \emph{physical models}, which combine physical and socio-economic concepts. We name three of them: microcanonical models~\cite{SUTTO2002,WYART2003}, models using Bose-Einstein statistics~\cite{BOTTA2003b,BOTTA2006} derived from an urn-and-ball scheme, and percolation models~\cite{FU2006} (see~\cite{CASTE2009} for a much deeper review).

Within the third category of \emph{stochastic models} is one of the oldest contributions to the literature, namely the \emph{Gibrat model} (1931)~\cite{gibra1931} (see also the reviews in~\cite{KALEC1945,SUTTO1997}). Many of the classical models assume Gibrat's Law, or some more sophisticated version of it~\cite{AXTEL2001,HART1996,HYMER1962,ISHIK2006b,MANSF1962,SIMON1958,SINGH1975}. Therefore, we describe it in some more detail. The model is based on the following assumptions:
\begin{enumerate}
\item \emph{Law of Proportionate Effect} or \emph{Gibrat's Law}: the absolute growth rate of a company is independent of its size, i.e.:
\begin{equation}\label{EqGibrat}
\frac{S_{t+\Delta t}}{S_t}=1+\epsilon_t,
\end{equation}
with $\Delta t$ the time period between measurements, $\epsilon_t$ an uncorrelated random noise, usually taken to be normally distributed with $\bar{\epsilon}=0$ and $\sigma_{\epsilon}\ll1$,
\item successive growth rates are uncorrelated in time,
\item companies do not interact.
\end{enumerate}

In order to measure growth, the central variable we look at is the growth rate, defined as $r(t,\Delta t)\equiv\log_{10}\frac{S_{t}}{S_{t-\Delta t}}$. The choice of $\Delta t$ (typically one year) is conditioned by the sampling frequency in the data sets. We use here $\Delta t=\text{1 year}$. Naming $S(t_i)\equiv S_i$, we write a rate in general as
\begin{equation}\label{EqGRdef2}
r_1\equiv\log_{10}\frac{S_1}{S_0},
\end{equation}
and call $S_0$ initial size (this term is not to be confused with the size at the initial time step $S_{\text{ini}}$; it is rather the size from which the growth rate is computed). We should also note that the statistical distributions depend on $\Delta t$~\cite{AMARA1997}. The size distribution is typically approximated by a log-normal. On the other hand, the growth rate in the Gibrat model follows a random-walk-like dynamics, and its distribution is normal. However, the literature agrees that a good way to describe at least the body of the growth-rate distribution for empirical data is through a Laplace (or ``tent-shaped'') distribution~\cite{AMARA1997b,AMARA1997,AMARA1998,BOTTA2003b}
\begin{equation}\label{EqLaplace}
p(r_1|S_0)=\frac{1}{\sqrt{2}\sigma_1(S_0)}\exp\left(-\frac{\sqrt{2}\left|r_1-\bar{r}_1(S_0)\right|}{\sigma_1(S_0)}\right),
\end{equation}
where $\bar{r}_1(S_0)$ is the mean value of the growth rates in the bin, and $\sigma_1(S_0)$ its standard deviation. This means that the Gibrat pattern is qualitatively very different from what one observes in real data. The growth-rate distributions for all initial-size bins are alike (due to Gibrat's Law), contrary to real observations, in particular regarding the decay of the fluctuations $\sigma_1(S_0)$ as $S_0$ increases. It is reported in the literature that a power law of the form $\sigma_1(S_0)\sim S_0^{-\beta}$ provides a good description for this decay~\cite{AMARA1997}. Moreover, the Laplace tails are ``fatter'' than Gaussian, i.e. extreme values have higher probability. This implies that large growth rates (both positive and negative) are more likely in reality than in the Gibrat model. In other words, organizational size changes very little most of the time, but it can occasionally also change dramatically.

Additionally, there is a subcategory of models called \emph{subunit models}, in which the size of a company is constructed as the sum of the contributions of internal subunits, e.g. different divisions. One well-known model is by Amaral et al.~\cite{AMARA1998}. A variation of this model is the transactional model~\cite{SCHWE2007}. Another variation represents groups as classes composed of subunits~\cite{FU2005,RICCA2008}. In the hierarchical tree model~\cite{AMARA1997c,AMARA1997} organizational hierarchy comes into play explicitly.

A final model in this category is based on additive replication in its general form. We call it \emph{general SAF model}, for Schwarzkopf, Axtell and Farmer who first proposed it~\cite{SCHWA2010}. A specific case of this model is the base for ours (see SAF model below). At each time step, each member of an organization is replaced by $x$ new ones, this last value taken from a replication distribution $p(x)$. There is a competition rule: the new element is either taken from another group with probability $\xi$, or created from scratch with probability $1-\xi$. The model is implemented on a social network, where vertices are individuals and edges are acquaintance relationships. The general SAF model is the only model among the reviewed literature that makes an explicit reference to a social network. This model follows our approach when it comes to designing a micro simulation model that generates a macro-level pattern through a defined mechanism.

The range of approaches from different disciplines shows the interdisciplinary nature of the phenomenon and the potential application of meaningful generative models to different scientific fields.

\section*{Methods}\label{SecModDef}
	\subsection*{SAF model}\label{SubSecStocMod}
\begin{figure}[!ht]
\begin{center}
\includegraphics[width=3.27in]{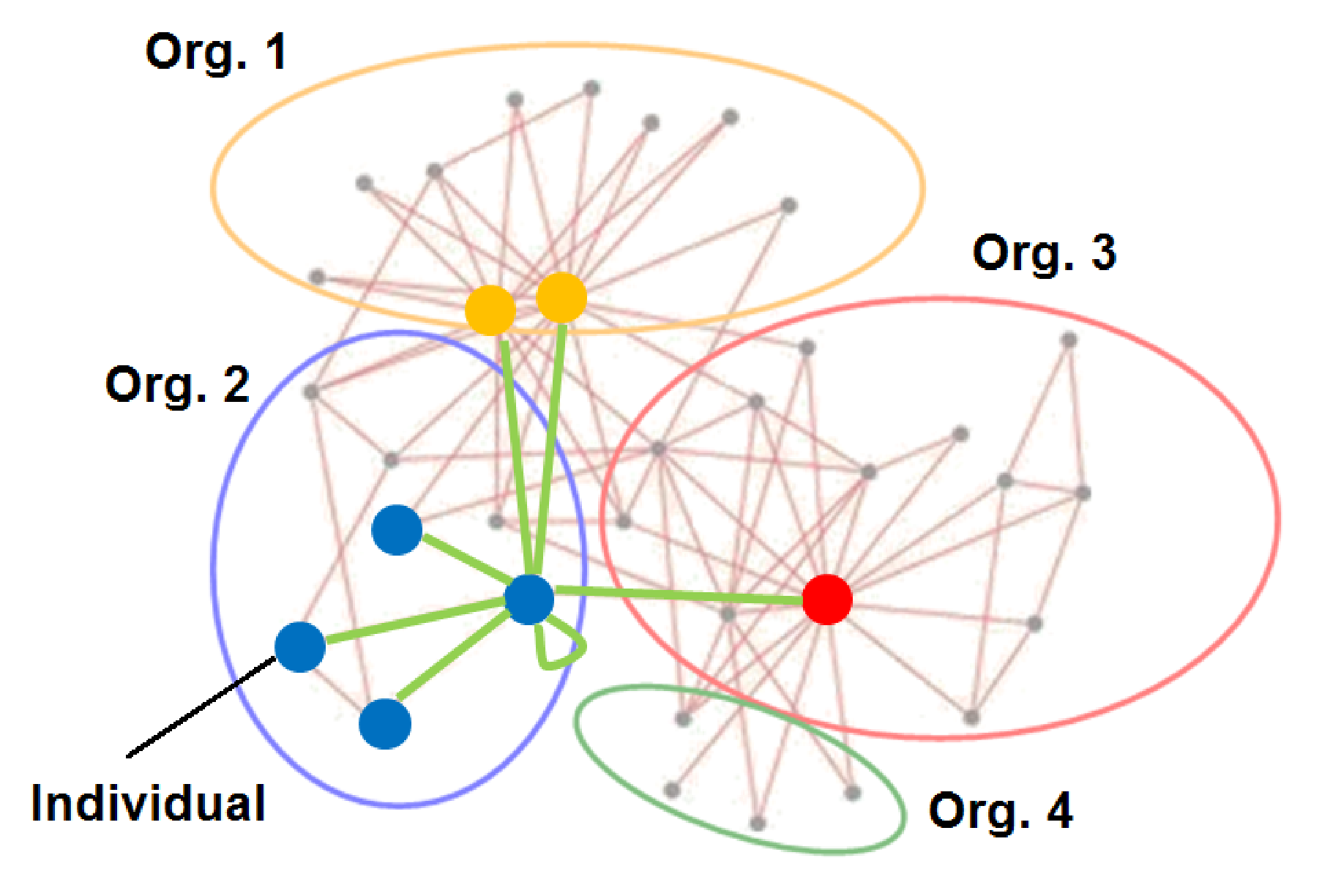}
\end{center}
\caption{
{\bf Setting for SAF model.}  The vertices of the network represent individuals, and the arcs are relations of contact influence between individuals. The network is static. An individual is for simplicity assumed to belong to one and only one organization. The size $S^{(\alpha)}(t)$ of organization $\alpha$ at time $t$ is the number of vertices belonging to that group at that time. At a certain time step, the probability for an individual to switch group depends on the group membership of its neighbourhood (highlighted in the figure). Note the loop on the individual to represent that the agent takes into account her own membership in the decision.}
\label{Fig_Syst}
\end{figure}

The dynamics by which people become members of an organization has many different aspects~\cite{WILSO2000}. One of them is influence through contacts in social networks, which we call \emph{contact influence}. The setting for the \emph{SAF model} consists then of a contact network. See the illustration in Fig.~\ref{Fig_Syst}. There are $N$ vertices representing individuals, and the arcs between them are links of contact influence. We follow~\cite{SCHWA2010} here. We work in the strong-competition limit ($\xi=0$): Each agent added to a group must be taken from another group. Consequently, the total number of agents $N$ is constant. The number of groups $G$ is fixed as well.

The only interaction we consider among agents is social influence, through edges in their contact networks. The network arc $i\rightarrow j$ meaning that agent $i$ is influenced by agent $j$. This simplification leaves out interactions coming from the formal (or informal) hierarchical structure of the organization. The underlying structure in the model is rather the social network of individual contacts, which we assume static over the time span of the problem.

The model variables are the sizes $S^{(\alpha)}(t)$ of each organization $\alpha$ at time $t$, with $\alpha\in\{1,2,\cdots,G\}$. The size of an organization at a certain time is the sum of the individuals in that group at that time. Previous research has shown that other size definitions ---for instance in terms of sales ---produce similar statistics (see e.g. Ref.~\cite{AMARA1997}).

Regarding the time evolution, at each time step $t$, a vertex is picked at random. The probability for vertex $i$ to switch to group $\eta$ we call switching probability, and is computed as
\begin{equation}\label{SAFswitchP}
P_{i,\eta}(t)=\frac{\tilde{k}_{i,\eta}(t)}{\sum_{\alpha}\tilde{k}_{i,\alpha}(t)},
\end{equation}
where $\tilde{k}_{i,\alpha}$ is the degree of vertex $i$ in group $\alpha$, counting its own group. That is to say, from the number of vertices that influence vertex $i$ (counting itself), how many belong to group $\alpha$. This rule conditions the decision to switch group on the group membership of network neighbours, and allows for the possibility to stay in the current organization.\\[0.25cm]

[Figure 1 here]\\

We impose an extra rule stating that no group can die out permanently. This is done to avoid that groups hit the absorbing state at size zero and thereby keeping the system in equilibrium. We implement the rule as follows: every Monte Carlo step a check is performed; if a group has zero size, a random vertex is switched to the empty group. A non-equilibrium version of the model is also possible. It would lead to different dynamics, with all system realizations going to a final one-group absorbing state. We tested this version as well, and found that the fat-tailed pattern is qualitatively reproduced. The implications are different, though. For instance, simulation time in the non-equilibrium version of the model could be translated more directly into some function of real time, while for an equilibrium model the association is not so direct.

One advantage with this model is that it has a clear sociological interpretation: the more close acquaintances a person has in a certain group, the more likely it is that the person will choose to become member of that group. The decision is mediated by a contact network, which puts an emphasis on the relational component of membership choice.

	\subsection*{Extended SAF model}\label{SubSecModExt}
So far we have proposed a model where an individual is more influenced to join a group the more acquaintances she has in that group at the time. Influence is exerted via the individual's contact neighbourhood. But social influence can be broader than that. Several models for social influence have been proposed, for example regarding culture~\cite{AXELR1997}, opinion formation~\cite{SZNAJ2005}, information sharing in groups~\cite{CARLE1991}, etc. Specifically, there is a contextual aspect of social influence. Different settings can entail different pressures towards homogeneity of opinions or membership~\cite{kuran1995}. We add one parameter to our model that represents the degree of this kind of social influence, which we call \emph{contextual influence}. The key element we want to incorporate is that influence in a certain setting is not a property of individual agents, but rather a property that affects all the members in the mentioned context. We assume, for simplicity, a uniform contextual influence. We model it through a parameter of \emph{contextual influence} $\delta$, with $0\leq\delta<\infty$. In the limit $\delta\rightarrow0$, the person does not feel any pressure to align herself with the neighbourhood. On the contrary, in the limit $\delta\rightarrow\infty$ the person acts solely based on the majority opinion in her surrounding neighbourhood.

We now define the \emph{extended SAF model}. The assumptions, parameters, and variables listed before are still valid. However, the time evolution is now governed by the following switching probability:
\begin{equation}\label{modSAFswitchP}
P_{i,\eta}(t)=\frac{\tilde{k}_{i,\eta}(t)^{\delta}}{\sum_{\alpha}\tilde{k}_{i,\alpha}(t)^{\delta}}.
\end{equation}

Setting $\delta=1$ recovers the first model. In the situation of low contextual influence ($\delta\rightarrow0$) a vertex can change its state at random, not being influenced by the groups of the contact vertices, while still retaining information on the possible groups she can choose to switch to. The system configuration tends thus to a random one. This can be thought as analogous to a high-temperature (disordered) situation in a physical system. In the situation of high contextual influence ($\delta\rightarrow\infty$) the vertex looks highly upon her contacts. Configurations where the vertex is not aligned with the majority of her neighbours become less and less likely, and the system tends to polarize itself in domains. This is the analogous of a low-temperature (ordered) situation, the difference being that a physical unit does not have global information about the total number of groups in the system.

\section*{Results}\label{SecResult}
	\subsection*{SAF model}\label{SubSecResSAF}
\begin{figure}[!ht]
\begin{center}
\includegraphics[width=3.27in]{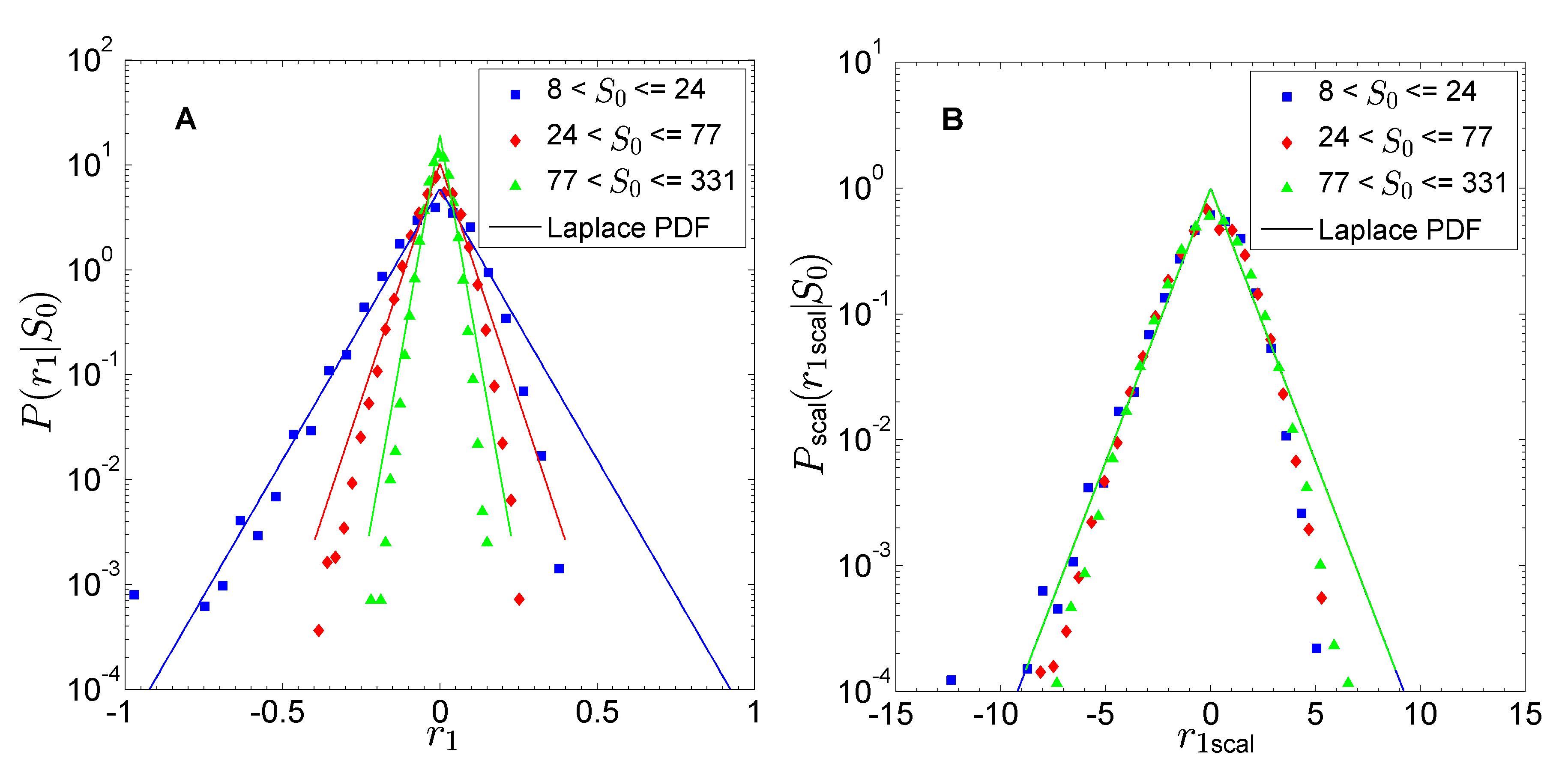}
\end{center}
\caption{
{\bf Growth-rate distribution for SAF model.}  (A) Growth-rate distribution. With $S_0$ the size at one year, and $S_1$ the size after one year, we define the growth rate $r_1\equiv\log_{10}(S_1/S_0)$. Here we plot the conditional PDF $p(r_1|S_0)$ to have a growth rate $r_1$ given an initial size $S_0$, in log-scale. The data is binned by initial-size ranges, and shown by organization type. We also plot a fit by MLE to the Laplace distribution in Eq.~(\ref{EqLaplace}). The overall fit is good, because that distribution carries most of its probabilistic mass in the body. (B) Same information, now in the scaled form of Eq.~(\ref{EqLaplaceScaled}). [Erd\H{o}s-R\'{e}nyi (ER) undirected network, $\left\langle k\right\rangle=10$, $N= 3,000$, $G=60$.]}
\label{Fig_SAFGRdist}
\end{figure}

We implement the SAF model in an Erd\H{o}s-R\'{e}nyi (ER) undirected network, by Monte Carlo simulation. The size distribution fits a log-normal distribution. The growth rate distribution for the basic case is shown in Fig.~\ref{Fig_SAFGRdist}A. All initial-size bins fit a Laplace distribution. The variance decay with an increase of $S_0$ is also verified. Additionally, Fig.~\ref{Fig_SAFGRdist}B plots the so-called scaled distributions (used in e.g.~\cite{AMARA1997b,AMARA1998,LEE1998}). Given the function in Eq.~(\ref{EqLaplace}), one can rescale the variables
\begin{align}\label{EqLaplaceScaled}
p_{\text{scal}}(r_1|S_0)=\exp&(-\left|r_{1,\text{scal}}\right|)\nonumber\\
r_{1,\text{scal}}&\equiv\frac{\sqrt{2}\left(r_1-\bar{r}_1(S_0)\right)}{\sigma_1(S_0)}\\
p_{\text{scal}}(r_1|S_0)&\equiv\sqrt{2}\sigma_1(S_0)p(r_1|S_0).\nonumber
\end{align}
Under this rescaling, the distributions for the different initial-size bins should collapse onto a single curve close to the Laplace distribution, as the figure shows.\\[0.25cm]

[Figure 2 here]\\

Looking at the simulated growth-rate distributions, the upper tail tends in general to underestimate the corresponding Laplace curve, while the lower tail follows it more closely. We interpret this as a consequence of working with constant $N$. In our model, membership growth in one organization is done at the expense of membership decline in the rest of the groups. This is reflected in less frequent positive growth rates. The fit is still good because the Laplace is a highly-peaked distribution, concentrating much of the mass around zero, so the main deviations represent a small fraction of the total deviation. The fact that real systems exhibit fatter tails on both sides could be due many factors, including the fact that growth-rate distributions could be a superposition of the distributions for different $N$s.

We then implement the simulation with different network properties, in order to see the impact of network structure on the observables.\\[0.25cm]

[Figure 3 here]\\

\begin{figure}[!ht]
\begin{center}
\includegraphics[width=3.27in]{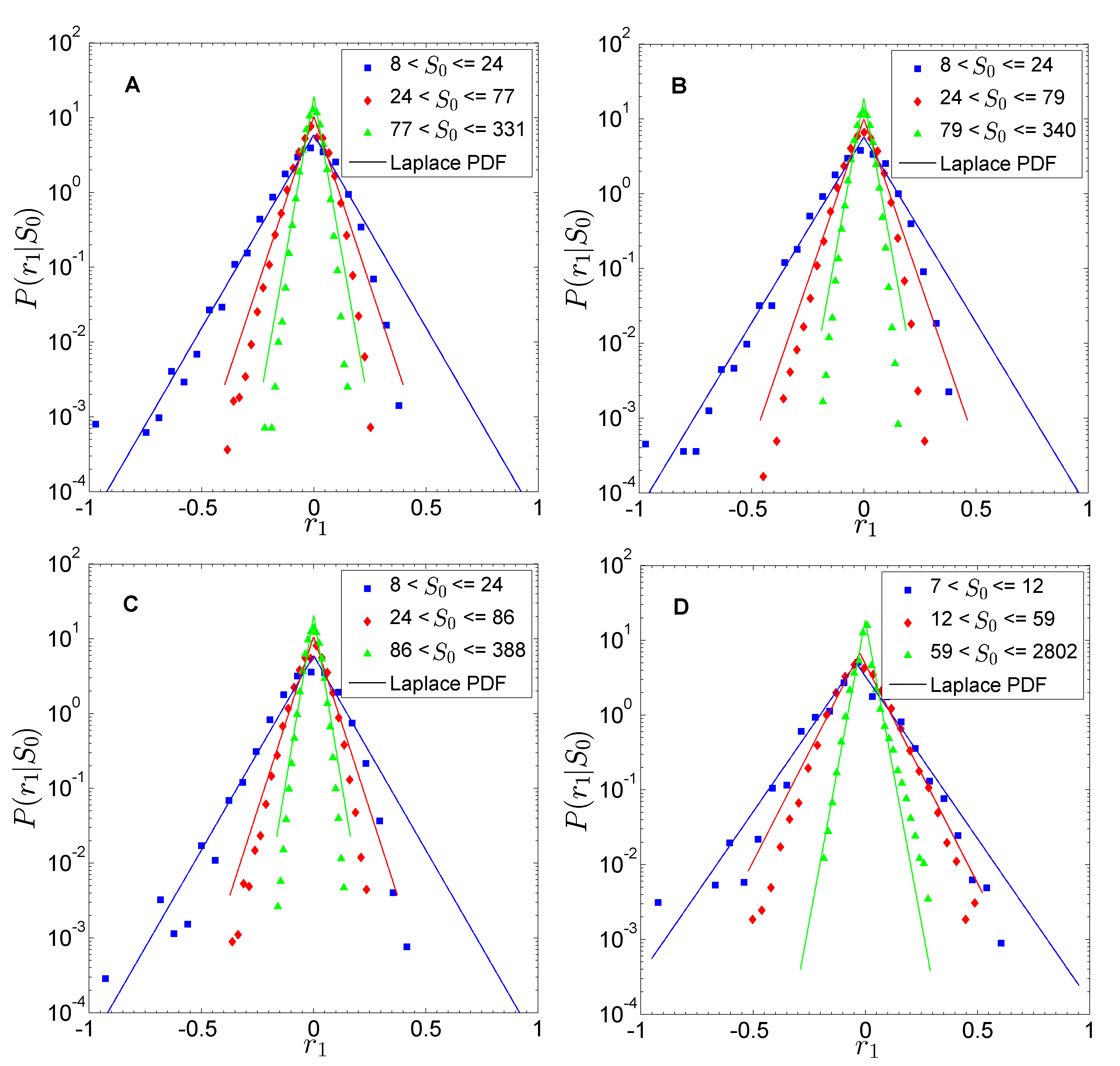}
\end{center}
\caption{
{\bf Changing network parameters in SAF model.}  We compare the original simulation in (A), with a different networks in (B--D). (B) ER directed network. The average behaviour does not change when changing the directionality. (C) Scale-free (SF) undirected network. When changing the type keeping the directionality, no qualitative change is observed. (D) SF directed network. The distributions get broader for all initial-size bins, and the third one concentrates most of the sizes. This reflects the nature of a scale-free directed network, with a few vertices being very influential/susceptible, and the majority having low degree. [$\left\langle k\right\rangle=10$, $N= 3,000$, $G=60$.]}
\label{Fig_NetParam}
\end{figure}

As a first change, we change the network from undirected to directed. We do so, keeping the mean degree constant, which implies that the number of influence arcs (incoming and outgoing) to a vertex remains on average unchanged. The reciprocity of each individual edge is lost, but the situation is still balanced on average, because the arc distribution along the network is random. That is to say, each agent on average influences and is influenced by the same number of alters. The comparison is illustrated by Fig.~\ref{Fig_NetParam}A--B. We can observe that the pattern is similar, both in terms of variance and of the ranges of initial-size bins.

Next, we change the network degree distribution from ER to scale-free (SF), again keeping the mean degree constant. The plots are shown in Fig.~\ref{Fig_NetParam}C--D. The undirected case is qualitatively similar, while the difference comes with the scale-free directed case. In the latter the distributions have larger variance in all initial-size bins, and more importantly, the higher initial-size bin comprises a much larger size range. The SF degree distribution appears to induce this behaviour, and our interpretation of this is as follows. In a SF network, by definition, there will be few ``popular'' vertices, and a lot of vertices with low popularity. As long as the influence is symmetric, the popular vertices drag its neighbours to their groups, but after a while, the equilibrium condition turns the tables--- the high degree of a few vertices just makes the process faster in certain moments. Imposing a directed network breaks the symmetry. In the directed case, some popular vertices are highly influential (they receive a lot of arcs, and consequently them changing their group impacts many other vertices) while other popular vertices are highly susceptible (they radiate many arcs, but a group change is not as influential). We understand that this asymmetry manifests itself on the dynamics, causing the growth-rate distributions to be broader, and a lot of peaks of high group size reflected in the higher $S_0$ range.

The fact that both ER and SF networks are able to generate the Laplace pattern can be interpreted in the light of the time-evolution rule of the model in Eq.~(\ref{SAFswitchP}). In effect, it is a rule implementing some kind of preferential attachment, with the probability to switch group becoming greater as the vertex degree increases.

	\subsection*{Extended SAF model}\label{SubSecResExtSAF}
\begin{figure}[!ht]
\begin{center}
\includegraphics[width=3.27in]{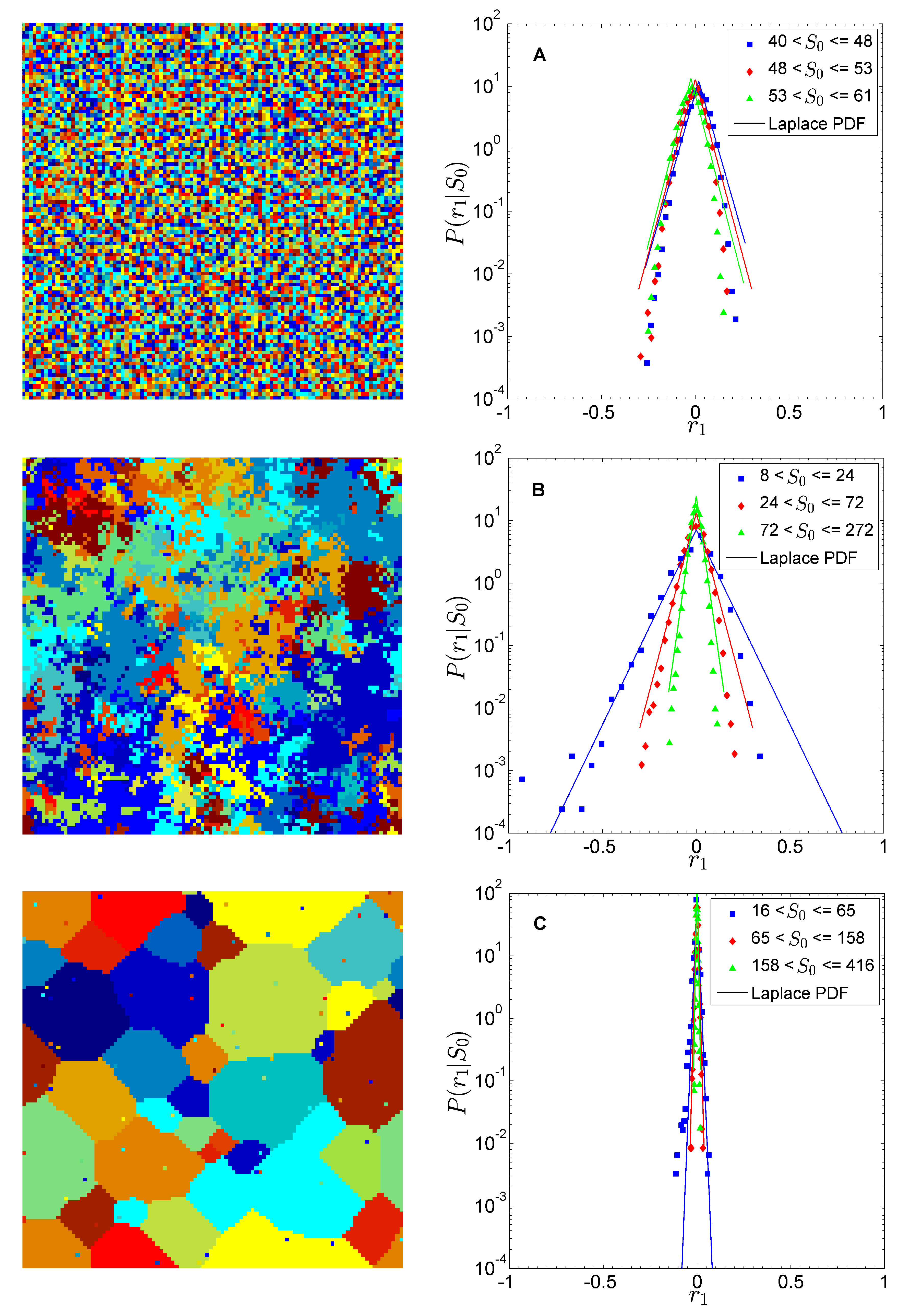}
\end{center}
\caption{
{\bf Group and growth-rate distributions for extended SAF model.}  We show the distributions on the right side, and on the left side snapshots of the last time step. As the degree of contextual influence $\delta$ increases, we observe how domains gradually appear. The higher the contextual influence, the more likely is that a vertex would align herself with her neighbours. (A) $\delta=0$. Low contextual influence, random behaviour similar to Gibrat model pattern. No domains exist. (B) $\delta=1$. Original SAF behaviour. Domains begin to appear. (C) $\delta=10$. High contextual influence. Presence of clear domains. [Square lattice, $\left\langle k\right\rangle=q=8$, $N= 10,000$, $G=200$.]}
\label{Fig_ModSAFdelta}
\end{figure}

Using the extended SAF model, we now test different contextual influences. The analysis is done on a square lattice to facilitate the visualization of domains of vertices belonging to the same organization. The lattice has periodic boundary conditions, and we use the Moore neighbourhood ($q=8$ nearest neighbours). In Fig.~\ref{Fig_ModSAFdelta}, we plot the group spacial distribution across the lattice, as well as the growth-rate distribution, for three values of $\delta$. The first situation, $\delta=0$, corresponds to a situation of low contextual influence--- the system has no clear domains, the organization assignment tending to a random uniform one. This is reflected in the growth-rate distribution with a pattern similar to the one encountered in the Gibrat model. The second situation, $\delta=1$, recovers the original SAF model. The third situation, $\delta=10$, corresponds to a situation of high contextual influence. There are clear domains where a few organizations absorb the majority of agents. This is reflected in the growth-rate distribution as a collapse of all distributions on highly-peaked curves close to $r_1=0$.\\[0.25cm]

[Figure 4 here]\\

\section*{Discussion}\label{SecDiscConc}
In our study we show how individual agents, having local information on membership alternatives and interacting with local simple rules through their social network, can generate fat-tailed macro patterns of organizational growth. In doing so, we have not assumed any institutional constrain or external perturbation. Rather, it is the internal dynamics of the interaction that bring the distribution about. Individual agents are subject to contact influence in their localized network neighbourhoods, but the aggregation of their individual membership decisions brings about unexpected macro-level outcomes. Sometimes, like in the case of large values of growth rates, the consequences at the macro level are quite extreme. This result is relevant for the design of policies and regulations, which are usually much more grounded in traditional approaches. Non-Gaussian patterns tend to challenge our worldview of how a lot of processes typically work.

The growth-rate fat-tailed pattern shows up in Erd\H{o}s-R\'{e}nyi, scale-free and square-lattice regular networks, both in terms of the Laplace distribution and in the decrease of the variance with $S_0$. This suggests that the mechanism driven by influence-based rules is more relevant to the pattern's qualitative replication than the details of network topology. Going back to our aims at the introduction, we find then that the scale-free character of the network is not a necessary condition to get a Laplace-like growth-rate distribution.

Looking at the results from the extended SAF model, the parameter region around $\delta=1$ provides a mechanism able to replicate the system's growth process features. When $\delta\rightarrow0$, the system has no clear group clusters, the organization assignment tending to a random one. On the contrary, when $\delta>1$, there are clear domains where a few organizations absorb the majority of agents. The intermediate situation best describes the real system's behaviour, and is modelled as a combination of contact and contextual influence. While the high-$\delta$ situation produces stable rich-get-richer dynamics, to the extent that single-group clusters do not break up, in the intermediate situation the rich-get-richer dynamics is no more stable. The situation around $\delta=1$ is also the only one where fluctuations are different for different initial sizes. This addresses our second aim, so that our model does not resort to the diversification argument to generate the decreasing variance with initial size.

The sociological interpretation of our results is that the transition zone where the real system exists is an intermediate situation, dominated by neither totally random behaviour nor totally compliant behaviour. It is possible to interpret this from the point of view of the information an agent has to have in order to act. One way to implement the SAF model ($\delta=1$) is to think that, at a given time step, an agent chooses a random link amongst her neighbours, and switches to that contact's group. Such an implementation means that, on average, each organization $\alpha$ will be picked with a probability equal to the corresponding extended vertex degree $\tilde{k}_{i,\alpha}$ of that agent. Each agent needs to know only the group membership of the contact she last encounters, making this situation a reasonable model of a dynamics where people successively meet contacts without any further information. The high-$\delta$ situation demands more information, since the agent should know the membership of all her contacts at a given time to be able to determine which is the majority membership. On the side of low contextual influence, choosing at random requires to know at least how many groups there are. So the intermediate case offers the agent a localized decision rule with minimal information requirements. We therefore get to a realistic model without invoking any argument of the real system being self-organized around a critical region. The parameter $\delta$ can thus be reinterpreted as a way to weight different choice strategies, and tuning it around $\delta=1$ recovers a case of bounded rationality~\cite{KAHNE2003}.

Therefore, we have to interpretations for $\delta$: the degree of contextual influence, and the tuning of membership choice strategy. The former is external to the individual, while the latter is internal. Both have an impact on the agent's behaviour, and both produce the aggregate pattern we observe empirically when tuned around the value for the SAF model. This suggests a duality between agent and social context, where the two views are consistent with the statistics we observe, and compatible with each other. We think this way of thinking exemplifies how to model one of the core issues in sociology, i.e., the interplay between individuality and social influence.

Further research should try to identify quantitatively how the statistical properties of the growth-rate distribution respond to systematic variations in both model and network parameters. For instance, in our explorations we found that the typical size of a group, given by $N/G$, seems to affect the distribution's variance. A significant model extension to consider would be to allow the system size $N$ to change in time. This variation would have to be implemented with care in this network approach, because the properties of the growing network should be monitored dynamically throughout the simulation. Other interesting extensions could be to incorporate community and hierarchical structure. The possibility to belong to more than one organization is another important point.

Another discussion concerns extensions to the parameter $\delta$. In this study we have introduced it as a parameter quantifying the effect of an agent's social context. Contextual influence enters as an exponent that weights the probability to switch group. In our formulation, the degree of contextual influence is uniform for all agents and constant in simulation time. There are relevant extensions to consider. For instance, one could assume that different types of organizations have particularities as to their social settings, and model this with a parameter that depends on organizational type. These different parameters can then be related to the growth-rate statistics. Additionally, this framework of analysis should be quite dependent on the size rage of the organizations under study, i.e. small voluntary-oriented organizations with local range have a setting where the social networks may dominate the dynamics, while large formal organizations have other structural elements in place so that a direct application of our model would not be advisable.

Finally, a better understanding of organizational growth processes could be applicable to other processes producing similar statistical features, from bird populations to financial and economic systems. This being said, one should still be careful in signaling the apparent universal presence of these common features as evidence of the systems belonging to the same class. On that line, it is reported that the exponent $\beta$ of the variance power-law relation, despite its value being similar for different systems, may not be universal. However, it is likely that different growth processes share similarities in terms of the underlying mechanisms driving them.

\section*{Acknowledgments}
HM and FL conceived the model; HM ran the simulations; HM, PH and FL analyzed the results; HM, PH and FL wrote the paper.


\begin{thebibliography}{xx}

\bibitem{AMARA1997b}
Amaral LAN, Buldyrev SV, Havlin S, Leschhorn H, Maass P, et~al. (1997) {Scaling Behavior in Economics I: Empirical Results for Company Growth}.
\newblock J Phys I (France) 7: 621-633.

\bibitem{AMARA1997c}
Buldyrev SV, Amaral LAN, Havlin S, Leschhorn H, Maass P, et~al. (1997) {Scaling Behavior in Economics II: Modeling of Company Growth}.
\newblock J Phys I (France) 7: 635-650.

\bibitem{AMARA1997}
Amaral LAN, Buldyrev SV, Havlin S, Maass P, Salinger MA, et~al. (1997) {Scaling behaviour in Economics: The problem of quantifying company growth}.
\newblock Physica A 244: 1-24.

\bibitem{AMARA1998}
Amaral LAN, Buldyrev SV, Havlin S, Salinger MA, Stanley HE (1998) {Power Law Scaling for a System of Interacting Units with Complex Internal Structure}.
\newblock Phys Rev Lett 80: 1385-1388.

\bibitem{BOTTA2003}
Bottazzi G, Secchi A (2003) {Common Properties and Sectoral Specificities in the Dynamics of U.S. Manufacturing Companies}.
\newblock Review of Industrial Organization 23: 217-232.

\bibitem{GUPTA2007}
Gupta H, Campanha JR, de~Aguiar DR, Queiroz GA, Raheja CG (2007) {Gradually truncated log-normal in USA publicly traded firm size distribution}.
\newblock Physica A 375: 643-650.

\bibitem{PLERO1999b}
Plerou V, Gopikrishnan P, Amaral LAN, Meyer M, Stanley HE (1999) {Scaling of the distribution of price fluctuations of individual companies}.
\newblock Phys Rev E 60: 6519-6529.

\bibitem{STANL1996}
Stanley MHR, Amaral LAN (1996) {Scaling behaviour in the growth of companies}.
\newblock Nature 379: 804-806.

\bibitem{BOTTA2001}
Bottazzi G (2001) {Firm Diversification and the Law of Proportionate Effect}.
\newblock LEM paper series.

\bibitem{BOTTA2007}
Bottazzi G (2007) {On the irreconcialability of Pareto and Gibrat Laws}.
\newblock LEM paper series.

\bibitem{BOTTA2002}
Bottazzi G, Secchi A (2002) {On the Laplace distribution of Firm Growth Rates}.
\newblock LEM paper series.

\bibitem{BOTTA2003b}
Bottazzi G, Secchi A (2003) {A stochastic model of firm growth}.
\newblock Physica A 324: 213-219.

\bibitem{BOTTA2006}
Bottazzi G, Secchi A (2006) {Explaining the Distribution of Firm Growth Rates}.
\newblock The RAND Journal of Economics 37: 235-256.

\bibitem{DOSI2005}
Dosi G (2005) {Statistical Regularities in the Evolution of Industries. A Guide through some Evidence and Challenges for the Theory}.
\newblock LEM paper series.

\bibitem{ISHIK2006}
Ishikawa A (2006) {Derivation of the distribution from extended Gibrat's law}.
\newblock Physica A 367: 425-434.

\bibitem{ISHIK2006b}
Ishikawa A (2006) {Pareto index induced from the scale of companies}.
\newblock Physica A 363: 367-376.

\bibitem{ISHIK2007}
Ishikawa A (2007) {The uniqueness of firm size distribution function from tent-shaped growth rate distribution}.
\newblock Physica A 383: 79-84.

\bibitem{HART1996}
Hart PE, Oulton N (1996) {Growth and Size of Firms}.
\newblock The Econ J 106: 1242-1252.

\bibitem{SINGH1975}
Singh A, Whittington G (1975) {The Size and Growth of Firms}.
\newblock Rev Econ Studies 42: 15-26.

\bibitem{lilje2001}
Liljeros F (2001) {The complexity of social organizing}.
\newblock Ph.D. thesis, Dept. of Sociology, Stockholm University, Stockholm, Sweden.

\bibitem{DEFAB2003}
De~Fabritiis G, Pammolli F, Riccaboni M (2003) {On size and growth of business firms}.
\newblock Physica A 324: 38-44.

\bibitem{FU2005}
Fu D, Pammolli F, Buldyrev SV, Riccaboni M, Matia K, et~al. (2005) {The Growth of Business Firms: Theoretical Framework and Empirical Evidence}.
\newblock PNAS 102: 18801-18806.

\bibitem{MATIA2004}
Matia K, Fu D, Buldyrev SV, Pammolli F, Riccaboni M, et~al. (2004) {Statistical properties of business firms structure and growth}.
\newblock Europhys Lett 67: 498-503.

\bibitem{FAGIO2007}
Fagiolo G, Napoletano M, Roventini A (2007) {How do output growth-rate distributions look like? Some cross-country, time-series evidence}.
\newblock Eur Phys J B 57: 205-211.

\bibitem{GAFFE2003}
Gaffeo E, Gallegati M, Palestrini A (2003) {On the size distribution of firms: Additional evidence from the G7 countries}.
\newblock Physica A 324: 117-123.

\bibitem{SCHWA2010}
Schwarzkopf Y, Axtell RL, Farmer JD (2010) {An explanation of universality in growth fluctuations}.
\newblock preprint SSRN ssrn.com/abstract=1597504.

\bibitem{CASTA2009}
Castaldi C, Dosi G (2009) {The patterns of output growth of firms and countries: Scale invariances and scale specificities}.
\newblock Empir Econ 37: 475-495.

\bibitem{LEE1998}
Lee Y, Amaral LAN, Canning D, Meyer M, Stanley HE (1998) {Universal Features in the Growth Dynamics of Complex Organizations}.
\newblock Phys Rev Lett 81: 3275-3278.

\bibitem{PODOB2008}
Podobnik B, Horvatic D, Pammolli F, Wang F, Stanley HE, et~al. (2008)
  {Size-dependent standard deviation for growth rates: Empirical results and theoretical modeling}.
\newblock Phys Rev E 77: 056102 1-8.

\bibitem{KEITT2002}
Keitt T, Amaral LAN, Buldyrev SV, Stanley HE (2002) {Scaling in the growth of geographically subdivided populations: invariant patterns from a
  continent-wide biological survey}.
\newblock Phil Trans R Soc Lond B 357: 627-633.

\bibitem{PLERO1999}
Plerou V, Amaral LAN, Gopikrishnan P, Meyer M, Stanley HE (1999) {Similarities between the growth dynamics of university research and of competitive
  economic activities}.
\newblock Nature 400: 433-437.

\bibitem{ijiri1977}
Ijiri Y, Simon HA (1977) {Skew distributions and the sizes of business firms}.
\newblock Amsterdam: North-Holland.

\bibitem{SIMON1964}
Simon HA (1964) {Comment: Firm Size and Rate of Growth}.
\newblock J Pol Econ 72: 81-82.

\bibitem{SIMON1958}
Simon HA, Bonini C (1958) {The Size Distribution of Business Firms}.
\newblock Amer Econom Rev 48: 607-617.

\bibitem{LUCAS1978}
Lucas R (1978) {On the Size Distribution of Business Firms}.
\newblock Bell J Econ 9: 508-523.

\bibitem{GUPTA2003}
Gupta H, Campanha JR (2003) {Firms growth dynamics, competition and power-law scaling}.
\newblock Physica A 323: 626-634.

\bibitem{JOVAN1982}
Jovanovic B (1982) {Selection and the Evolution of Industry}.
\newblock Econometrica 50: 649-670.

\bibitem{SUTTO2002}
Sutton J (2002) {The variance of firm growth rates: the ‘scaling’ puzzle}.
\newblock Physica A 312: 577-590.

\bibitem{WYART2003}
Wyart M, Bouchaud JP (2003) {Statistical models for company growth}.
\newblock Physica A 326: 241-255.

\bibitem{CASTE2009}
Castellano C, Fortunato S, Loreto V (2009) {Statistical physics of social dynamics}.
\newblock Rev Mod Phys 81: 591-646.

\bibitem{FU2006}
Fu D, Buldyrev SV, Salinger MA, Stanley HE (2006) {Percolation model for growth rates of aggregates and its application for business firm growth}.
\newblock Phys Rev E 74: 036118 1-7.

\bibitem{gibra1931}
Gibrat R (1931) {Les In\'{e}galit\'{e}s \'{e}conomiques}.
\newblock Paris: Recueil Sirey.

\bibitem{KALEC1945}
Kalecki M (1945) {On the Gibrat Distribution}.
\newblock Econometrica 13: 161-170.

\bibitem{SUTTO1997}
Sutton J (1997) {Gibrat's Legacy}.
\newblock J Econ Literature 35: 40-59.

\bibitem{AXTEL2001}
Axtell RL (2001) {Zipf Distribution of U.S. Firm Sizes}.
\newblock Science, New Series 293: 1818-1820.

\bibitem{HYMER1962}
Hymer S, Pashigian P (1962) {Firm Size and Rate of Growth}.
\newblock J Pol Econ 70: 556-569.

\bibitem{MANSF1962}
Mansfield E (1962) {Entry, Gibrat's Law, Innovation, and the Growth of Firms}.
\newblock Amer Econom Rev 52: 1023-1051.

\bibitem{SCHWE2007}
Schweiger A, Buldyrev SV, Stanley HE (2007) {A transactional theory of fluctuations in company size}.
\newblock preprint arXiv physics/0703023v1.

\bibitem{RICCA2008}
Riccaboni M, Pammolli F, Buldyrev SV, Ponta L, Stanley HE (2008) {The Size Variance Relationship of Business Firm Growth Rates}.
\newblock PNAS 105: 19595-19600.

\bibitem{COLEM1986}
Coleman JS (1986) {Social Theory, Social Research, and a Theory of Action}.
\newblock Amer J Sociology 91: 1309-1335.

\bibitem{hedst2005}
Hedstr\"{o}m P (2005) {Dissecting the social: on the principles of analytical sociology}.
\newblock Cambridge: Cambridge University Press.

\bibitem{GRANO1985}
Granovetter M (1985) {Economic Action and Social Structure: The Problem of Embeddedness}.
\newblock Amer J Sociology 91: 481-510.

\bibitem{WILSO2000}
Wilson J (2000) {Volunteering}.
\newblock Annu Rev Sociol 26: 215-240.

\bibitem{AXELR1997}
Axelrod R (1997) {The Dissemination of Culture: A Model with Local Convergence and Global Polarization}.
\newblock J Conflict Resolut 41: 203-226.

\bibitem{SZNAJ2005}
Sznajd-Weron K (2005) {Sznajd model and its applications}.
\newblock Acta Phys Pol B 36: 2537-2547.

\bibitem{CARLE1991}
Carley KM (1991) {A Theory of Group Stability}.
\newblock American Sociological Review 56: 331-354.

\bibitem{kuran1995}
Kuran T (1995) {Private truths, public lies: the social consequences of   preference falsification}. \newblock Cambridge: Harvard University Press.

\bibitem{KAHNE2003}
Kahneman, D (2003) {Maps of Bounded Rationality: Psychology for Behavioral Economics}.
\newblock The American Economic Review 93: 1449-1475.

\end{thebibliography}
%

\section*{Figure Legends}

\end{document}